**Multisensory learning recruits visual neurons into an olfactory memory engram**

Abbreviated title: Multisensory learning expands a memory engram

Zeynep Okray[1,*], Nils Otto[1,2], Anna A. Cook[1], Clifford Talbot[1], Ashwin Miriyala[1], Martín Klappenbach[1], Ciara Stern[1], Kieran Desmond[1], Paola Vargas-Gutierrez[1], and Scott Waddell[1,*]

[1]Centre for Neural Circuits & Behaviour, University of Oxford, Oxford, OX1 3TA, United Kingdom

[2]Present address: Institute of Anatomy and Molecular Neurobiology, Westfälische Wilhelms-University, 48149 Münster, Germany

*Corresponding authors: Scott Waddell (scott.waddell@cncb.ox.ac.uk) and Zeynep Okray (zeynep.okray@cncb.ox.ac.uk)

Lead contact: Scott Waddell

**Abstract**

Associating multiple sensory cues with a single experience or object is a fundamental process that improves object recognition and memory performance. However, neural mechanisms that bind sensory features during learning and augment memory expression are unknown. Here we demonstrate multisensory appetitive and aversive memory in *Drosophila*. Combining colours and odours improved memory performance, even when each sensory modality was tested alone. Temporal control of neuronal function revealed visually-selective mushroom body Kenyon Cells (KCs) to be required for enhancement of visual and olfactory memory recall after multisensory training. Synapse-level connectomics suggests that valence-relevant dopaminergic reinforcement could permit the KC-spanning serotonergic DPM neurons to bridge between previously 'modality-selective' KC streams. Consistent with this model, DPM transmission is uniquely required during multisensory memory formation and for enhanced expression of olfactory memory afterwards. In addition, signalling via the DopR1 dopamine receptor is required in APL neurons, suggesting that reinforcing dopamine could locally release GABA-ergic inhibition to permit bridging microcircuits to function. Cross-modal binding thereby expands the KCs representing the olfactory memory engram into those representing the colour. We propose that broadening of the engram improves memory performance after multisensory learning and permits a single sensory feature to retrieve the memory of the multimodal experience.

Life is a rich multisensory experience for most animals. As a result, nervous systems have evolved to use multisensory representations of objects, scenes and events to most effectively guide behaviour[1]. It is widely appreciated that multisensory learning improves memory performance, from children in the classroom to rodents and insects in controlled laboratory experiments[2–5]. Moreover, an apparently universal and unexplained feature of multisensory learning is that it improves subsequent memory performance even for the separate unisensory components[3,5]. Studies in humans and mammals have suggested that multisensory learning benefits from interactions between modality specific cortices that were co-active during training, and that individual senses can reactivate both areas at testing[3,6–9]. In addition, cells in different brain regions respond to multiple sensory cues and the proportions/numbers change after multisensory learning[1,10–12]. However, we currently lack detailed mechanistic understanding of how multisensory learning converts neurons from being modality selective to multimodal, and how enhanced multisensory and unisensory memory performance could be supported by such a process.

In *Drosophila*, unique populations of mushroom body (MB) Kenyon Cells (KCs) receive predominant and anatomically segregated dendritic input from olfactory or visual projection neurons (as well as local visual interneurons) and their axons project as parallel streams into the MB lobes. There, successive compartments of each KC's axonal arbor are intersected by the presynapses of dopaminergic neurons (DANs) that convey the reinforcing effects of appetitive or aversive stimuli[13,14]. Reinforcing dopamine depresses synapses between active KCs and the compartment-restricted dendrites of downstream mushroom body output neurons (MBONs) to code valence-specific memories[15–17].

**Multisensory learning improves memory**

To study multisensory learning in *Drosophila* we adapted the olfactory T-maze[18] so that colours and odours can be presented together (Figure. 1a). Food-deprived flies were trained by presenting them with a colour and/or odour (conditioned stimulus minus, CS−), followed by another colour and/or odour (conditioned stimulus plus, CS+) paired with sugar reward (Figure. 1a-b). When trained and tested with only colours [Visual learning, V], flies did not show significant learned preference for the previously sugar paired colour (Figure. 1b-d; Figure 1 – figure supplement 1a, 24h memory). However, combining colours with odours [Congruent protocol, C] produced robust and long-lasting memory, which was significantly enhanced over that formed by training with only odours [Olfactory learning, O] (Figure 1b-d, and Figure 1 - figure supplement 1a). If the colour and odour pairs were reversed between training and testing [Incongruent protocol, I] (Figure 1b-d; Figure 1 - figure supplement 1a,b), no memory enhancement was observed. Furthermore, memory enhancement was not apparent if the same colour was presented with CS- and CS+ odours during training and testing (Figure 1 figure supplement 1c, C v Csc). The memory enhancing effect of multisensory learning therefore requires a learned relationship between specific colour and odour combinations.

To further investigate multimodal memory enhancement, we restricted presentation of multisensory cues to either training or testing. Multisensory training improved memory retrieval even when testing used unisensory cues [Olfactory Retrieval, OR; Visual Retrieval, VR] (Figure 1e,f). In contrast, presenting multisensory stimuli during testing did not improve performance in olfactory trained flies [Multisensory Retrieval, MSR] (Figure 1 - figure supplement 1d). Moreover, the greatest improvement in performance was observed when multisensory stimuli were used during training and testing (Figure 1 - figure supplement 1e). Therefore, multisensory training enhances memory performance for the individual colour and odour memory components, and congruence of colour and odour information between training and testing further improves performance. Although our experiments and elsewhere [5,19] imply that flies distinguish green and blue colours, we do not discount a contribution of hue and luminance.

**Olfactory retrieval requires visual KCs**

Dendrites of the numerically larger populations of olfactory KCs within the αβ, α′β′ and γ lobes (ie. γ main, γm KCs) occupy the main calyces of the MB, whereas the relatively small populations of αβ posterior (αβp) and γ dorsal (γd) KCs receive predominantly visual information via dendrites in the accessory calyces[14]. γd KCs were previously implicated in colour learning[19,20]. We tested roles for visual γd and αβp KCs in multisensory learning and memory, using cell-specific expression of a UAS-*Shibire*$^{ts1}$ (*Shi*$^{ts1}$) transgene which encodes a dominant temperature-sensitive dynamin[21]. At temperatures >30 ˚C, *Shi*$^{ts1}$ blocks membrane recycling and thus impairs synaptic transmission. Function can be restored by returning flies to <29˚C. Blocking output from γd and αβp KCs during testing at 6h abolished visual enhancement of performance in the congruent protocol, and removed the interference of incongruent color cues. In both instances memory performance was similar to that of flies tested with odours alone (Figure 2a-d, Figure 2 - figure supplement 1a-f). These results suggest activity in γd and αβp KCs represents the visual component of multisensory memory (see also Figure 2 - figure supplement 1g-h). Surprisingly, blocking γd KC (but not αβp KCs) output during testing also impaired performance for odour-only memory retrieval in multisensory trained flies (Figure 2e-f), despite having no effect on memory retrieval in flies trained with only odours[22] (Figure 2 - figure supplement 1a,d). This unexpected result led us to hypothesize that multisensory learning might expand the representation of odours to include 'visual' γd KCs.

**DPM neurons are positioned to bridge KC sensory streams**

Anatomy of the MB network suggests two possible ways to confer odour responsiveness to γd KC axons: via KC-KC synapses or neurons positioned to bridge the different KC streams. We queried the anatomical feasibility of these routes using the complete MB connectome of a single adult female fly 'hemibrain' electron microscope (EM) volume[32,33]. Although most (562 of 590) γm KCs make synapses with γd KCs, number and placement of these connections does not support every γd KC to receive γm input in every γ lobe compartment. In addition, KC-KC connections were reported to suppress activity in neighboring KCs[34]. We therefore moved to identify neurons that could bridge different KC sensory streams, and studied the fine anatomy of the γ lobe innervation of the potentially excitatory serotonergic Dorsal Paired Medial (DPM) neuron in the hemibrain EM volume. DPM neurons send separate branches that densely innervate the vertical and horizontal lobes and distal peduncle of the MB, where they are both pre- and post-synaptic to KCs[35–37]. Ultrastructure of DPM neuron projections in the γ lobe revealed two branches within γ1 and other ventral and dorsal branches passing through the γ2 - γ5 compartments (Figure 3a). Positions of DPM neuron synapses onto γd KCs follow the γd KC axon bundle as it winds around the γ lobe from ventral in γ1 to dorsal in γ5 compartment (Figure 3a).

Annotating a dendrogram of DPM neurites (Figure 3b) with γ lobe compartment boundaries (based on DAN connectivity), synapses from γm KCs and those to γd KCs, showed that unique branches of the DPM neuron can provide compartment-specific microcircuit bridges between γm and γd KCs. DPM neurons can also bridge γd to γm KCs (Figure 3c). The large GABAergic anterior paired lateral (APL)[38] neuron was found to make synapses along DPM branches in the γ lobe (Figure 3b,c), suggesting DPM bridging can be regulated by local inhibition. Importantly, the APL neuron receives many DAN inputs within each compartment and can therefore also be regulated with region specificity[39] (Figure 3d), to potentially release specific DPM branches from APL inhibition.

**APL and DPM neurons bridge sensory streams in multisensory memory**

We challenged a putative microcircuit bridge model by independently manipulating APL and DPM neurons. Expression in APL neurons of the Dop2R dopamine receptor has been linked to aversive learning[40] and transciptional profiling suggests APL neurons also express DopEcR receptor[41]. Both of these dopamine receptors are inhibitory[42,43]. We therefore used *tub*P-GAL80[ts 44] to temporally-restrict transgenic RNAi expression in APL neurons to test a role of these receptors in multisensory learning. Knocking down Dop2R in adult APL neurons abolished multisensory enhancement of olfactory retrieval performance (Figure 4a, Figure 4 - figure supplement 1a). A mild defect was also observed for odour memory following olfactory appetitive conditioning, however the difference was only significant to one control (Figure 4b, Figure 4 - supplement 1b). In contrast, DopEcR RNAi had no effect in either experiment (Figure 4 -figure supplement 1c,d). These results are consistent with inhibition of APL neurons by reinforcing dopamine, to allow recruitment of γd KCs into the olfactory memory engram during multisensory learning.

We tested a role of the serotonergic DPM neurons in expanding the olfactory engram to include visually responsive KCs using expression of UAS-$Shi^{ts1}$. Temporally restricting transmission from DPM neurons either during acquisition (Figure 4c, Figure 4 - supplement 1e) or retrieval (Figure 4d; Figure 4 - supplement 1e) significantly impaired the olfactory retrieval of multisensory memory. Blocking DPM neuron output also impaired retrieval of visual memory after multisensory learning (Figure 4 - supplement 1f,g). These same manipulations had no effect on odour memory after unisensory olfactory learning (Figure 4 - supplement 1h), consistent with prior work[45]. These data are consistent with DPM neuron output being required during learning to bind together simultaneously active KC streams, whereas DPM neuron output during memory retrieval provides the connection for odour-driven γm KCs to activate the relevant γd KCs.

If multisensory learning establishes DPM neuron microcircuit bridges between olfactory γm and visual γd KCs, output from the γm KCs should be required during multisensory learning, like that of DPM neurons. We therefore used UAS-$Shi^{ts1}$ to test whether γm was required during learning for enhancement of olfactory and visual memory retrieval after multisensory training. Whereas blocking γ KCs during multisensory learning significantly impaired both olfactory and visual retrieval (Figure 4e, f, Figure 4 – figure supplement 1i, j), it did not alter olfactory learning (Figure 4g), as in prior studies[46–48]. A requirement for γm KC output for multisensory memory is therefore consistent with the DPM bridge model, and suggests that multisensory learning involves different plasticity rules to that of unisensory olfactory memory[16,49].

For DPM neuron-released serotonin (5-hydroxytryptamine, 5-HT) to mediate a bridge from γm to γd KCs, there should be a requirement for KC expressed 5-HT receptors. Since 5-HT can exert excitatory effects through 5-HT2A and 5-HT7 type receptors[51], we used RNAi to knockdown these receptors in γd KCs. Reducing 5-HT2A but not 5-HT7 or 5-HT2B receptor expression impaired olfactory memory performance after multisensory training (Figure 4h) but not olfactory training (Figure 4 – figure supplement 1k). Taken together the anatomical and behavioural genetic data lead us to propose that reinforcer-evoked compartment specific dopamine releases APL-mediated inhibition. This could facilitate the same reinforcing dopamine to induce KC-DPM and DPM-KC plasticity that forms excitatory serotonergic DPM microcircuit bridges between the relevant γm and γd KCs activated by specific odour-colour combinations (Figure 4i).

**Engram expansion benefits new learning**

An expansion of the CS+ odour representation into a particular segment of the γd axons after multisensory training might be expected to facilitate subsequent learning with the same odour, if the next DAN teaching signal intersects the expanded KC representation. We reasoned that if DANs direct the recruitment of γd axons to become odour activated, an aversive learning event which requires dopamine release from the PPL1-γ1pedc DANs that innervate the most proximal γ1 lobe compartment[31] should confer odour-responsiveness onto all downstream γd axon segments from γ1 to γ5 (Figure 5a). In contrast, reward learning which requires dopamine from PAM-γ4 and PAM-γ5 DANs, might only make these terminal segments of the γd KCs odour responsive (Figure 5b). We first confirmed that multisensory colour and odour aversive (electric shock) training produced memory enhancement for both combined and individual cues, similar to that observed by others [5], and after appetitive training (Figure 5 – figure supplement 1a-f), and that γd KCs were also required for odour memory enhancement following aversive multisensory training (Figure 5 – figure supplement 1g-l).

We then tested the axonal recruitment model by sequentially training flies with either an aversive (dopamine in γ1 and γ2) or appetitive (dopamine in γ4 and γ5) multisensory protocol followed by unisensory odour-reward or odour-punishment learning (Figure 5c,d). Prior multisensory aversive training significantly enhanced subsequent odour-reward learning (Figure 5c). However, no enhancement was apparent if aversive odour learning followed multisensory appetitive learning (Figure 5d). These data are consistent with a multisensory training-dependent expansion of the CS+ odour representation being included in the next CS+ odour memory engram when appropriate γd axon segments have become CS+ odour-activated. Larger colour and odour memory engrams provide a mechanism for how odour and colour memory performance is enhanced following multisensory training (Figure 1e,f), and explains why odour memory retrieval in this context acquires a requirement for γd KC output (Figure 2e).

**Discussion**

Our study describes a neural mechanism in *Drosophila* through which multisensory learning improves subsequent memory performance, even for individual sensory cues. A single training trial with visual cues could only generate robust memory performance if they were combined with odours during training, similar to visual rhythm perception learning in humans, which requires accompanying auditory information[52]. We propose that multisensory learning binds together information from temporally contingent odours and colours within axons of mushroom body γ KCs, via serotonergic DPM neurons, whose activity also defines the coincidence time window[53]. For this to occur, we reason that learning-driven binding converts axons of visually (presumably colour) selective KCs to also become responsive to the temporally contingent trained odour. We assume the reverse may also apply, that axons of olfactory selective KCs become activated by the temporally contingent trained colour. While predominant dendritic input defines γm KCs as being olfactory and γd as being visual, the anatomy of DPM neurons, the positions of reinforcing DANs, and sequential learning experiments suggest that segments of their axons may become multimodal after multisensory learning. This result suggests that γ KCs are a likely substrate where other temporally contingent sensory information can be integrated with that of explicit sensory cues[54,55].

Although our experiments mostly focused on odour activated γm KCs recruiting colour γd KCs via DPM microcircuits, the observed behavioural enhancement of visual memory following multisensory learning and the reciprocal connectivity of DPM neurons suggests that DPM neurons could mediate a reverse bridge, and even cross modal interactions between other sensory streams. In so doing, multisensory learning uses DPM neurons to link KCs that are responsive to each temporally contingent sensory cue and expands representations of each cue into that of the other. Such a cross-modal expansion would allow multisensory experience to be efficiently retrieved by combined cues and by each individually. As a result, trained flies can evoke a memory of a visual experience with the learned odour, and memory of an odour with the learned colour. These findings provide a neural mechanism through which the fly achieves a conceptual equivalent of hippocampus-dependent pattern completion in mammals, where partial cues can retrieve a more complete memory representation[56]. Interestingly, human patients with schizophrenia and autism exhibit deficits in multisensory integration[57] and these conditions have been linked to serotonergic dysfunction and 5-HT2A receptors[58]. Our work here suggests that inappropriate routing of multisensory percepts may contribute to these conditions. Moreover, the excitatory 5-HT2A receptors that mediate multisensory binding, are the major targets of hallucinogenic drugs[59].

**Figure Legends**

**Figure 1** with one supplement. **Multisensory learning enhances memory performance.**
**a.** *Left*, Apparatus for multisensory training and testing. *Right*, experimental timeline. **b.** Protocols. Green and blue squares represent colours, light and dark gray squares represent OCT and MCH odours. Visual (V) learning: colours used as CS+ and CS-. Olfactory (O) learning: odours used as CS+ and CS-. Congruent (C) protocol: colours+odours combined as CS+ and CS-. Same colour+odour combinations used for training and testing. Incongruent (I) protocol: colours+odours combined as CS+ and CS- but combinations were switched between training and testing. Olfactory retrieval (OR): colour+odour combinations used for training but only odours for testing. Visual retrieval (VR): colour+odour combinations used for training but only colours for testing. **c** and **d**. *Top*, training and testing timelines. *Bottom*, immediate (**c**) and 6 h (**d**) memory for (V), (O), (C) and (I) protocols. **e** and **f**. *Top*, timelines. *Bottom*, multisensory training with colours+odours tested immediatly (**e**) and 6 h (**f**) after training for each individual modality. Asterisks denote significant difference ($P < 0.05$). Data presented as mean ± standard error of mean (SEM). Individual data points displayed as dots correspond to independent experiments. Groups compared using one-way ANOVA with Tukey's test (**c**, **d**) and unpaired two-sided *t*-test (**e**,**f**), exact $P$ values and comparisons are given in Supplementary Table 1. N values for each experiment are: **c**, n=8 for V and n=10 for O, C and I; **d**, n=10; **e**, n=10; **f**, n=14 for O and OR and n=10 for V and VR.

**Figure 1 – figure supplement 1. Memory performance is most robust when colour and odour combinations are consistent during acquisition and retrieval.**
**a.** *Top left*, multisensory protocols. Green and blue squares represent colours, light and dark gray squares represent OCT and MCH odours. Visual (V) learning: colours used as CS+ and CS-. Olfactory (O) learning: odours used as CS+ and CS-. Congruent (C) protocol: colours+odours were combined as CS+ and CS-. Same colour+odour combinations used during training and testing. Incongruent (I) protocol: colours+odours were combined as CS+ and CS- but combinations were switched between training and testing. *Bottom left*, training and testing timeline. *Right*, 24 h memory performance for V, O, C and I protocols. Combining colours+odours in the congruent (C) protocol enhanced 24 h performance, compared to that obtained with unisensory V or O learning. Incongruent (I) pairing of colours and odours abolished the multisensory enhancement of 24 h memory. **B.** *Left*, training and testing timeline. *Middle*, multisensory protocols. *Right*, immediate memory performance. Flies showed a significantly higher memory following the C than the I protocol. **C.** *Left*, training and testing timeline. *Middle*, multisensory protocols: C protocol as described above; Congruent protocol using the same colour (C-sc) combined with different odours as CS+ and CS- during training and testing. *Right*, the C protocol using distinct colour+odour combinations for CS+ vs CS- resulted in higher immediate memory performance than the C-sc protocol using the same colour with both odours. **d.** *Left*, training and testing timeline. *Middle,* multisensory protocols: Olfactory (O) learning as described above*;* Multisensory Retrieval (MSR): odours were CS+ and CS- during training and these same odours were combined with different colours during testing. *Right*, immediate memory performance evoked by MSR was not significantly reduced to that following for Olfactory learning and retrieval. **e.** *Left*, training and testing timeline. *Middle,* multisensory protocols: Congruent (C) protocol as described above; Odour Retrieval (OR): colours+odours were CS+ and CS- during training and only odours were used during testing. *Right*, flies trained with multisensory stimuli performed better if they were tested with congruent multisensory stimuli compared to only one modality (in this case odour). Asterisks denote significant differences (*P < 0.05*). Data presented as mean ± standard error of mean (SEM). Individual data points displayed as dots correspond to independent experiments. Groups compared using one-way ANOVA with Tukey's test (**a**), unpaired two-sided *t*-test (**b, c**, **e**) and unpaired two-sided Mann–Whitney *U*-test (**d**), exact *P* values and comparisons are given in Supplementary Table 2. N values for each experiment are: **a**, **b**, **e**, n=10; **c**, n=8; **d**, n=10 for OR and n=8 for MSR.

**Figure 2** with one supplement. **Enhanced performance following multisensory learning requires visually-responsive γd and αβ$_p$ Kenyon Cells.**
**a.** *Top left*, schematic of γ dorsal [γd] KCs. *Bottom left*, timeline with temperature shifting (dashed line). *Right*, blocking output of γd KCs during testing using MB607B-GAL4; UAS-*Shi*[ts1] in the congruent protocol. **b**. Blocking γd KC output during testing in the incongruent protocol. **c.** *Top left*, schematic of αβp KCs. *Bottom left*, timeline with temperature shifting. *Right*, blocking αβp KC output during testing using c708a-GAL4; UAS-*Shi*[ts1] in the congruent protocol. **d**. Blocking αβp KC output during in the incongruent protocol. **e** and **f**. *Top*, timeline with temperature shifting. *Bottom,* blocking γd (**e**) and αβp (**f**) KCs output during testing of Olfactory Retrieval of multisensory memory. Asterisks denote significant difference (*P < 0.05*). Data presented as mean±SEM. Individual data points displayed as dots correspond to independent experiments. All groups compared using one-way ANOVA with Tukey's test, exact *P* values and comparisons are given in Supplementary Table 1. N values for each experiment are: **a**, **b**, **d**, **f**, n=12; **c**, **e**, n=10.

**Figure 2 – figure supplement 1. Constitutively blocking γd or αβp KC output impairs the visual component of multisensory memories.**
**a.** *Top left*, schematic of γd KCs. *Bottom left*, training and testing timeline with constant restrictive temperature (dashed line). *Right*, 6 h memory performance following Olfactory learning is unchanged when γd KCs are blocked through the experiment using MB607B-GAL4; UAS-*Shi*[ts1] **b** and **c**. Blocking γd KCs throughout the experiment significantly impaired memory in the Congruent protocol (**b**). The release from the interference effect of the Incongruent protocol did not reach significance (**c**). **d**. *Top left*, schematic of αβp KCs. *Bottom left*, training and testing timeline with constant restrictive temperature (dashed line). *Right*, blocking αβp KC output throughout the experiment using c708a-GAL4; UAS-*Shi*[ts1] did not impair 6 h Olfactory learning. **e** and **f**. Memory performance for Congruent (**e**) and Incongruent (**f**) protocols changed significantly when αβp KCs were blocked during the experiment. **g** and **h**. *Top*, training and testing timeline with temperature shifting (dashed line). Blocking αβp (**g**) and γd (**h**) KCs impaired memory retrieved with Visual cues. **h**. *Top left*, schematic of γd KCs. **i**. *Top*, training and testing timeline with constant permissive temperature (dashed line), applies to i-l. **i-l** Memory performance in MB607B-GAL4; UAS-*Shi*[ts1] flies following Congruent (**i**), Incongruent (**j**), Olfactory Retrieval (**k**) and Visual Retrieval (**l**) protocols was not affected when training and testing was performed at 23 °C. **m**. *Top left*, schematic of αβp KCs. *Bottom left*, training and testing timeline with constant permissive temperature (dashed line). **m-o** Memory performance in c708a-GAL4; UAS-*Shi*[ts1] flies after Congruent (**m**), Incongruent (**n**), and Visual Retrieval (**o**) protocols was not affected when training and testing was at 23 °C. Asterisks denote significant differences ($P < 0.05$). Data presented as mean ± standard error of mean (SEM). Individual data points displayed as dots correspond to independent experiments. Groups compared using one-way ANOVA with Tukey's test (**a-g, i-o**), and Kruskal–Wallis *H*-test with Dunn's test (**h**), exact *P* values and comparisons are given in Supplementary Table 2. N values for each experiment are: **a-c**, **g**, **h**, n=12; **d**, n=12 for c708a and n=10 for all other groups, **f**, n=10; **i-o**, n=8.

**Figure 3. DPM and APL connectivity to KCs allows for multisensory stimulus binding.**
**a**. *Left*, 3D representation of MB (light gray) with DPM neuron γ lobe neurites (teal). DPM trifurcates (asterisk) into dorsal (dark teal), ventral (light teal) and γ1 compartment (teal) branches. Neurites shaded by Strahler order and twigs with Strahler order <1 were pruned. *Right*, detail of the γ lobe with γ1-γ5 compartment borders (dashed lines). DPM presynapses to γd KCs (yellow spheres) co-localize on dorsal branch in γ5. Input synapses from γm KCs (gray spheres) to DPM locate throughout ventral and dorsal branches. In γ5, 450 of 585 γm KCs made synapses with the DPM dorsal branch, where 89 of 98 γd KCs also receive DPM input. APL neuron inputs (magenta spheres) localize along both DPM branches. **b**. 2D dendrogram projection of DPM neurites (shades of teal). The dorsal branch of the horizontal lobe's γ2-γ5 compartments are dark teal, the rest of the γ lobe projections are mid teal and those in the other lobes are lighter teal. Note: all but γ lobe neurites are downsized for visibility.

Neurites with Strahler order <1 are pruned and connectivity is shown accordingly. Projections in the γ1 compartment are marked and split in the γ2-5 compartments into dorsal and ventral branches. Marking is according to DAN connectivity (shaded areas). Synapses (spheres) are only marked in the γ lobe compartments. γ1 compartment is marked and γ2-5 are split between dorsal and ventral DPM neuron branches. Inputs from γm KCs (gray spheres) and outputs to γd KCs (yellow spheres) co-localize on compartment-specific branches. Inhibitory inputs from APL (magenta spheres) are distributed across DPM neurites. **c.** A 2-dimensional dendrogram projection of DPM neuron neurites as in **b** but with the reverse polarity of inputs from γd and KCs and γm KCs shown. Inputs from γd KCs (yellow spheres) and outputs to γm KCs (gray spheres) also colocalize on compartment specific branches of the DPM neurons. APL inputs (magenta) are distributed across both dorsal and ventral branches and concentrated around the branching point at the base of the vertical lobes in the α'1 compartment (asterisk). Note: Some putative parts of the dorsal γ lobe branch could not be allocated to a compartment due to lack of DAN input. One branch that bears γ KC synapses (hashtag) was identified as the β' lobe branch that enters the MB close to the midline. It is likely that other DPM neurite tips that have γ KC synapses are artifacts where healing has merged free-floating γ lobe branchlets to DPM neurites innervating the other lobes. **d.** *Left*, 3D representation of MB (light gray) with APL neuron γ lobe neurites (purple). Neurites shaded by Strahler order and twigs with Strahler order <1 were pruned. A 2-dimensional dendrogram projection of APL neuron neurites (magenta). Note: Neurites with Strahler order <1 were pruned and connectivity is shown accordingly. Compare with (**a**). PPL1-γ1pedc and PAM-γ5 DAN input synapses (red and green spheres respectively) are marked in the γ1 and γ5 compartments (shaded areas). Synapses to DPM neurons (teal spheres) colocalize with DAN inputs in the γ lobe.

**Figure 4** with one supplement. **DPM and APL neurons are required for multisensory memories.**
**a.** *Top*, Anatomy of APL neuron. Training and testing timelines with constant restrictive temperature (dashed line). *Bottom*, testing 6 h Olfactory Retrieval memory performance after (**a**) multisensory appetitive training, and (**b**) unisensory Olfactory learning in flies with adult-restricted RNAi knockdown of Dop2R in APL neurons using *tub*PGAL80[ts]; VT43924-GAL4.2. **c.** *Top*, DPM neuron schematic. Timeline with temperature shifting. *Bottom*, blocking DPM neuron output with VT64246-GAL4; UAS-*Shi*[ts1] during training (**c**) or testing (**d**) 6 h Olfactory Retrieval performance. **e-g.** *Top left*, schematic of γ KCs. *Top*, training and testing timeline with temperature shifting (dashed line). *Bottom*, blocking output of γ KCs during multisensory training using MB009B-GAL4; UAS-*Shi*[ts1] and testing 6 h Olfactory Retrieval (**e**) or Visual Retrieval (**f**), or (**g**) during olfactory training and testing 6 h olfactory memory performance. **h.**

Timeline. RNAi knockdown of 5-HT2A, 5-HT2B or 5-HT7 receptors in γd KCs with MB607B-GAL4 and testing of Olfactory Retrieval 6 h after multisensory training. Data represented as mean±SEM. Individual data points displayed as dots. Asterisks denote significant difference ($P < 0.05$). Groups compared using one-way ANOVA with Tukey's test (**a-h**), exact *P* values and comparisons are given in Supplementary Table 1. N values for each experiment are: **a**, **b**, n=12; **c**, n=8 for *Shi* and n=9 for other groups; **d**, n=10; **e-g**, n=8 flies; **h**, n=10. **i.** Model of DPM microcircuit bridging of odour and colour specific KCs following multisensory learning.

**Figure 4 – figure supplement 1. Roles for DPM and APL neurons in multisensory memories.**

**a** and **b**. *Top*, Anatomy of APL neuron. Training and testing timelines at constant permissive temperature (dashed line). No defect was observed for Olfactory Retrieval of appetitive multisensory memory (**a**) or of memory following Olfactory learning (**b**) in flies with repressed expression of Dop2R RNAi in APL neurons using *tub*PGAL80$^{ts}$; VT43924-GAL4.2. **c** and **d.** 6h Olfactory Retrieval of appetitive multisensory memory (**d**) and of memory following Olfactory learning (**b**) was not affected by adult-restricted RNAi knockdown of DopEcR in APL neurons using *tub*PGAL80$^{ts}$; VT43924-GAL4.2. **e.** *Top*, Anatomy of DPM neuron. *Top*, training and testing timeline. *Bottom*, Multisensory memory performance evoked by Olfactory Retrieval of VT64246-GAL4; UAS-*Shi*$^{ts1}$ flies was unaffected when flies were trained and tested at permissive 23˚C. **f.** *Top*, training and testing timeline with temperature shifting (dashed line). *Bottom*, blocking output from DPM neurons during testing impaired the Visual Retrieval of multisensory memory. **g.** *Top*, training and testing timeline. *Bottom*, Multisensory memory performance evoked by Visual Retrieval of VT64246-GAL4; UAS-*Shi*$^{ts1}$ flies was unaffected when flies were trained and tested at permissive 23˚C. **h.** *Top*, training and testing timeline with temperature shifting (dashed line). *Bottom*, blocking output from DPM neurons during unisensory training and testing did not affect Olfactory learning performance. **i.** *Top left*, anatomy of γ KCs. *Top*, training and testing timelines. *Bottom*, Multisensory memory performance evoked by Olfactory Retrieval (**i**) or Visual Retrieval (**j**) of MB009B-GAL4; UAS-*Shi*$^{ts1}$ flies was unaffected when trained and tested at 23˚C. **k.** *Top left*, Anatomy of γd KCs. *Bottom left*, training and testing timeline. *Right*, 6h Olfactory learning performance was unaffected in flies with RNAi knockdown of 5-HT2A with MB607B-GAL4. Asterisks denote significant differences ($P < 0.05$). Data presented as mean ± standard error of mean (SEM). Individual data points displayed as dots correspond to independent experiments. All groups compared using one-way ANOVA with Tukey's test, exact *P* values and comparisons are given in Supplementary Table 2. N values for each experiment are: **a**, **b**, n=18; **c**, **d**, n=10; **e-g**, n=8; **h**, n=12; **i-k**, n=8.

**Figure 5** with one supplement. **Engram expansion benefits new learning**

**a.** MB model for appetitive multisensory colour+odour training followed by unisensory odour or colour testing. γm KCs receive dendritic olfactory input and γd visual input. Both γKC types project axons through γ1- γ5 compartments of the MB γ lobe. Appetitive training (*left*) engages reward DANs (green) innervating γ4 and γ5, whose released dopamine encodes learning by depressing synapses[49] between odour-activated KCs and avoidance-directing Mushroom Body Output Neurons (MBONs; not illustrated)[14]. Dopamine signalling during multisensory learning also binds γm and γd KC activity in γ4-5 compartments. During future unisensory odour testing (*middle*), odour excites specific γm KCs (thick gray), which in turn activate γd axons in γ4-5 compartments (gray dashed lines to yellow). Reverse γd-mediated activation of γm KCs occurs with unisensory colour testing (*right*). **b.** MB model for aversive multisensory training followed by odour testing. Aversive multisensory training (*left*) engages punishment DANs (red) that depress synapses[16,17] between γm and γd KCs and approach-directing γ1 and γ2 MBONs[16,17] (not shown) while also binding γd and γm KC activity in these compartments. Unisensory odour testing (*right*) excites specific γm KCs which activate γd axons from γ1 forward. **c** and **d**. Prior aversive multisensory learning enhances future appetitive but not aversive odour learning. *Left*, protocols. Starved flies were divided into: Group I, aversive multisensory training, and Group II, aversive unisensory odour training. 3 h later both groups were trained with odours and sugar reward (using the same CS+/CS- odours as for initial training) and tested immediately afterwards. *Right*, memory performance. **c.** Group I initially trained with multisensory aversive protocol performed better than Group II initially aversively trained with only odours. **d.** Group I initially trained with multisensory appetitive protocol did not outperform Group II initially trained with only odours. Asterisks denote significant differences ($P < 0.05$). Data presented as mean±SEM. Individual data points displayed as dots and correspond to independent experiments. Groups compared using unpaired two-sided *t*-test (**c**,**d**), exact $P$ values and comparisons are given in Supplementary Table 1. N values for each experiment are: **c**, n=12; **d**, n=8.

**Figure 5 – figure supplement 1. Multisensory aversive learning enhances memory for the combined and individual odour and colour cues.**

**a.** *Top*, aversive training and testing timeline. *Bottom*, multisensory experimental conditions. Green and blue squares represent colours, light and dark gray represent OCT and MCH odours. Visual (V) learning; Olfactory (O) learning; Congruent (C) protocol; Incongruent (I) protocol; Olfactory Retrieval (OR); Visual Retrieval (VR). **b-d.** *Top*, training and testing timelines. *Bottom*, aversive memory with C protocol was significantly increased to that with I protocol both immediately (**b**) and 3 h (**c**) after training. Flies in the C protocol outperformed those tested 3 h after Olfactory (O) Learning. Only the C protocol generated significant 24 h memory performance (**d**. **e** and **f**. *Top*, training and testing timelines. *Bottom,* when tested immediately (**e**) multisensory memory retrieved with colours (VR) was markedly better than that following unisensory Visual learning (V). At 3 h (**f**) multisensory trained flies performed significantly better when their memory was retrieved with only Olfactory (OR) or Visual (VR) cues than flies trained with unisensory Olfactory (O) or Visual (V) learning. **g**. *Top left*, schematic of γd KCs. *Bottom left*, training and testing timeline with temperature shifting (dashed line). **g-i.** Blocking output of γd KCs during testing using MB607B-GAL4; UAS-*Shi*$^{ts1}$ alters 3 h memory performance in the Congruent (**g**), Incongruent (**h**) and Olfactory Retrieval (**I**) protocols. **j.** *Top left*, schematic of γd KCs. *Bottom left*, training and testing timeline with constant permissive temperature (dashed line). **j-l** Memory performance in MB607B-GAL4; UAS-*Shi*$^{ts1}$ flies after Congruent (**j**), Incongruent (**k**), and Olfactory Retrieval (**l**) protocols was not affected when flies were trained and tested at 23 °C. Asterisks denote significant differences (*P < 0.05*). Data presented as mean ± standard error of mean (SEM). Individual data points displayed as dots correspond to independent experiments. **m** and **n**. Groups compared using one-way ANOVA with Tukey's test (**b-d**, **g-l**), unpaired two-sided *t*-test (**e**, **f**), exact *P* values and comparisons are given in Supplementary Table2. N values for each experiment are: **b**, **g**, **i-l**, n=8; **c**, **e**, **f**, **h**, n=10; **d**, n=12.

**Methods**

**Fly strains** All *Drosophila melanogaster* strains were reared at 25 °C and 40-50% humidity, except where noted, on standard cornmeal-agar food (100 g l$^{-1}$ anhydrous d-glucose, 47.27 g l$^{-1}$ organic maize flour, 25 g l$^{-1}$ autolyzed yeast, 7.18 g l$^{-1}$ agar, 12.18 g Tegosept dissolved in 8.36 ml absolute ethanol, per liter of fly food) in 12:12 h light:dark cycle. Canton-S flies were used as wild-type (WT) and originated from William Quinn's laboratory (Massachusetts Institute of Technology, Cambridge, MA, USA). The following GAL4 lines were used in the behaviour experiments: MB607B-GAL4[60,61], MB009B-GAL4[60,61], c708a-GAL4[62], VT43924-GAL4.2[63], VT64246-GAL4[64]. Temperature controlled blocking of neuronal output was achieved by expressing the UAS-*Shi*$^{ts1}$ [65] transgene under the control of the MB607B-GAL4[60,61], MB009B-GAL4[60,61], c708a-GAL4[62] and VT64246-GAL4[64] drivers. For RNAi knockdown experiments involving APL, *tub*P-GAL80$^{ts}$ [66], VT43924-GAL4.2[63] flies were crossed with UAS-Dop2R RNAi[67] and UAS-DopEcR RNAi (VDRC ID 103494) flies. The same driver line was crossed to wild-type flies and the RNAi background strain (VDRC ID 60100), as controls. For RNAi knockdown experiments involving γd KCs, MB607B-GAL4[60,61] flies were crossed with UAS-5-HT2a RNAi (BDSC 31882), UAS-5-HT2b RNAi (BDSC 60488) and UAS-5-HT7 RNAi (BDSC 27273) flies. The same driver line was crossed to the RNAi background strain (BDSC 36304), as controls. We used both male and female flies for all experiments.

**Behavioural experiments** Male flies from the GAL4 lines were crossed to UAS-*Shi*$^{ts1}$ virgin females, except for experiments involving c708a-GAL4, where UAS-*Shi*$^{ts1}$ males were crossed to c708a-GAL4 virgin females. For heterozygous controls, GAL4 or UAS-*Shi*$^{ts1}$ flies were crossed to WT flies. In RNAi experiments GAL4 or RNAi flies were crossed to the appropriate RNAi background strains or WT flies, respectively. All flies were raised at 25 °C, except where noted below for manipulation of RNAi expression. Populations of 2–8 day-old flies were used in all experiments.

For appetitive conditioning experiments, 80-100 flies were placed in a 25 ml vial containing 1% agar (as a water source) and a 20 × 60 mm piece of filter paper for 19–22 h before training and were kept starved for the entire experiment, except when assaying 24 h memory where flies were fed for 30 min after training then returned to starvation vials until testing. For aversive conditioning experiments, 80-100 flies were placed in a vial containing standard food and a piece of filter paper for 14–22 h before behavioural experiments.

For experiments involving neuronal blocking with UAS-*Shi*$^{ts1}$, a schematic of the timeline of temperature shifting is provided in each figure. For *Shi*$^{ts1}$ experiments, flies were transferred to restrictive 33 °C 30 min prior to training and/or testing. For RNAi experiments involving *tub*P-GAL80$^{ts}$; VT43924-GAL4.2, flies were raised at 18 °C and shifted to 29 °C after eclosion to induce RNAi expression for 3 d before the behavioural experiments. The flies remained at 29 °C for the duration of the experiments.

All behavioural experiments were conducted using a standard T-maze that was modified to allow simultaneous delivery of colour and odour stimuli. The T-maze which is made from translucent plastic was covered in opaque blackout film to minimise interference between the visual stimuli when they were used in parallel. Odours were 4-methylcyclohexanol (MCH) and 3-octanol (OCT) diluted in mineral oil (at ≈1:10$^{-3}$ dilution). Colours were provided by light-emitting diodes (LEDs); green LEDs with wavelength 530 +/- 10nm (ProLight Opto, PM2E-3LGE-SD) and blue LEDs with wavelength 465 +/- 10nm (ProLight Opto, PM2B-3LDE-SD). Four LEDs were assembled in a circuit built onto a heat sink and were mounted securely on top of the odour delivery tubes. The intensities of the LEDs were adjusted so that naïve flies showed no phototactic preference between the illuminated T-maze arms. Visual stimuli were presented in the same manner and same intensity for both training and testing. For appetitive experiments the testing tubes were lined with filter paper and for aversive experiments the testing tubes were lined with non-electrified shock grids. Experiments were performed in an environmental chamber set to the desired temperature and 55-65% relative humidity. Flies were handled prior to training and testing under overhead red light.

Appetitive conditioning was performed essentially as previously described[68]. Briefly, flies were exposed for 2 min to stimuli Y ($Y_{Colour}$ and/or $Y_{Odour}$) without reinforcement in a tube with dry filter paper (the Conditioned Stimulus minus, CS-), 30 s of clean air, then 2 min with stimuli X ($X_{Colour}$ and/or $X_{Odour}$) presented with 5.8 M sucrose dried on a filter paper (the Conditioned Stimulus plus, CS+). For aversive olfactory conditioning[69,70], flies recieved 1 min exposure to stimuli X ($X_{Colour}$ and/or $X_{Odour}$) paired with twelve 90 V electric shocks at 5 s intervals (CS+), 45 s of clean air, followed by 1 min exposure to stimuli Y ($Y_{Colour}$ and/or $Y_{Odour}$) without reinforcement (CS-). Electric shocks were delivered using a Grass S48 Square Pulse Stimulator (Grass Technology). Shock grids are those previously described[71] and consist of interleaved copper rows printed on transparent Mylar film which allows coloured light to pass through.

Memory performance was assessed by testing flies for their preference between the CS- and the CS+ colours and/or odours for 2 min. Odour testing was performed in darkness. The flies in each arm were collected and transferred to polystyrene tubes (Falcon 14mL Round Bottom

Polypropylene Test Tube with Cap). Tubes with flies were frozen at -20°C and flies were then removed and manually counted.

Performance Indices were calculated as the number of flies in the CS+ arm minus the number in the CS- arm, divided by the total number of flies. For all behavioural experiments, a single sample, or *n*, represents the average Performance Index from two independent groups of flies trained with the reciprocal colour/odour combinations as CS+ and CS-. The total n for each experiment was acquired over 3 different training sessions on different days.

Six behavioural protocols were used:
1) Visual (V) Learning: colours ($X_{Colour}$ and $Y_{Colour}$) were used as CS+ and CS-.
2) Olfactory (O) Learning: odours ($X_{Odour}$ and $Y_{Odour}$) were used as CS+ and CS-.
3) Congruent (C) protocol: colours and odours were combined ($X_{Colour}$ + $X_{Odour}$ and $Y_{Colour}$ + $Y_{Odour}$) as CS+ and CS-. The same colours and odours combinations were used during training and testing.
4) Incongruent (I) protocol: colour and odours stimulus contingencies were switched between training ($X_{Colour}$ + $Y_{Odour}$ and $Y_{Colour}$ + $X_{Odour}$) and testing ($X_{Colour}$ + $X_{Odour}$ and $Y_{Colour}$ + $Y_{Odour}$).
The V and O protocols are unisensory, whereas C and I are multisensory.
5) Olfactory retrieval (OR): flies were trained as in the Congruent (C) protocol , but only odours ($X_{Odour}$ and $Y_{Odour}$) were presented as the choice at test.
6) Visual retrieval (VR): flies were trained as in the Congruent (C) protocol , but only colours ($X_{Colour}$ and $Y_{Colour}$) were presented as the choice at test.
The sequential learning experiments depicted in Figure 5c and 5d used aversive or appetitive Congruent (C) multisensory training followed by unisensory appetitive or aversive Olfactory (O) learning, then testing using olfactory retrieval (OR).

**Statistical Analysis** Statistical analyses were performed in GraphPad Prism. All behavioural data were analyzed with an unpaired two-sided *t*-test, Mann-Whitney *U*-test, one-way ANOVA or Kruskal-Wallis *H*-test followed by a posthoc Tukey's, Dunnet's or Dunn's multiple comparisons test. No statistical methods were used to predetermine sample size. Sample sizes were similar to other publications in the field.

**Blinding, Randomization** The experiments were randomized with appropriate controls present in each independent experiment. All genotypes tested and analyzed were self-blinded to the experimenter.

**Neuroanatomy, Connectivity, Dendrograms** Neuromorphological calculations and connectivity analyses were performed, and dendrograms calculated & plotted, with scripts based on NAVis 1.2.1 library functions in Python 3.8.8 (https://pypi.org/project/navis/; https://github.com/navis-org/navis)[86] and data from the *Drosophila* hemibrain (v.1.2.1) (https://neuprint.janelia.org)[87,88]. All neuronal skeletons were healed (navis.heal_skeleton (method = "ALL", max_dist = "100 nanometer", min_size = 10)), rerooted (navis.reroot_skeleton (x.soma)) and strongly down sampled with conserved connectors (navis.downsample_neuron( downsampling_factor = 1000, preserve_nodes = 'connectors' )). 3D representations of neurons shaded by Strahler order were generated with navis.plot2d (method='3d', shade_by='strahler_index'), after pruning twigs with Strahler order ≤1 (navis.prune_by_strahler()). Where applicable only branches in specific volumes were considered (navis.in_volume()). Volumes were obtained from neuprint (v.1.2.1) with fetch_roi(). Connectivity was analyzed using unpruned neurons and with compartment specificity (navis.in_volume()).

Custom scripts based on navis.plot_flat() were used to generate dendrograms of DPM and APL neurons with twigs of Strahler order ≤1 pruned. MB compartment boundaries were defined by connectivity to DANs of the respective compartments. Branches outside the γ-lobe were downsized manually to increase visibility of γ-lobe compartments. Synapses are filtered by in_volume() and displayed on branches with Strahler order >1. Connectivity statistics are based on unpruned neurons and synapses between neurons were obtained with R based natverse:: neuprint_get_synapses() (https://natverse.org)[86] scripts and processed with custom scripts in Python.

**Code availability.**

Code used for Methods section "Neuroanatomy, Connectivity, Dendrograms" is publicly available at https://pypi.org/project/navis/, https://github.com/navis-org/navis and https://natverse.org.

**Data availability.**

Supplementary Table 1 and Supplementary Table 2 and data supporting the findings of this study are available at https://github.com/CNCBWaddellTmaze/multisensory-memory-Drosophila. Dataset of *Drosophila* hemibrain used for Methods section "Neuroanatomy, Connectivity, Dendrograms" is publicly available at https://neuprint.janelia.org.

**Acknowledgments** We thank R. Brain, K. delos Santos, R. Busby, M. Moreno Gasulla, J-P. Moszynski, F. Woods for technical assistance. P. Jacob contributed to behavioural experiments. We thank G. Rubin, FlyLight, B. Dickson, A. Lin and the Vienna Drosophila

Resource Center and the Bloomington Stock Center for flies. Z.O and A.A.C were funded by EMBO Long-term postdoctoral fellowships (ALTF 311-2017 and ALTF 1300-2024). P.V-G. was funded by the Sloane Robinson/Clarendon Scholarship given by the University's Clarendon Fund scheme and Keble College and from the Consejo Nacional de Ciencia y Tecnología (CONACYT). S.W. was funded by a Wellcome Principal Research Fellowship (200846), a Wellcome Discovery Award (225192), an ERC Advanced Grant (789274), and Wellcome Collaborative Awards (203261 and 209235).

**Author contributions** Designed research Z.O., C.T., A.A.C., A.M., M.K., S.W., Performed research Z.O., N.O., A.A.C., A.M., M.K., C.S., K.D., P.V.G., Analyzed data Z.O., N.O., C.S., K.D., A.A.C., A.M., M.K., S.W. Resources S.W. Writing S.W., Z.O., N.O., A.A.C., Supervision S.W., Funding Acquisition S.W., Z.O., A.A.C.

**Conflict of Interest** The authors declare no competing financial interests.

**Correspondence and requests for materials** should be addressed to S.W and/or Z.O.

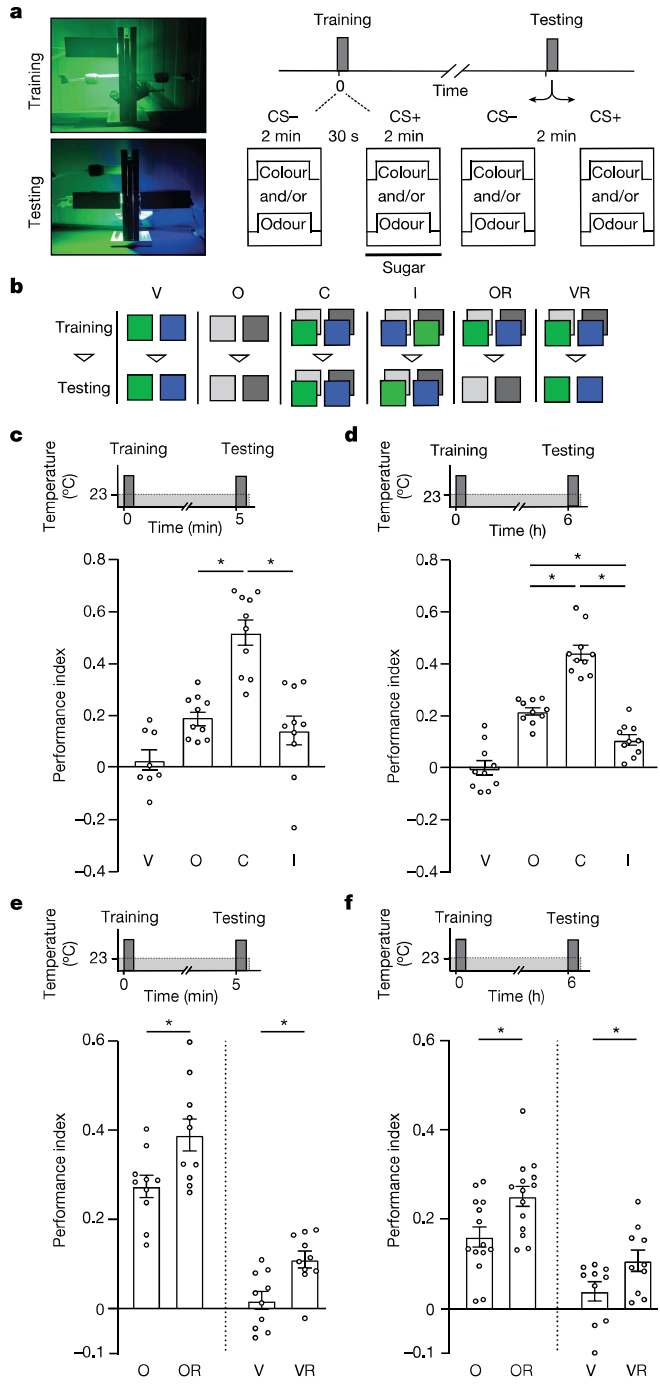

Figure 1 Supplement 1

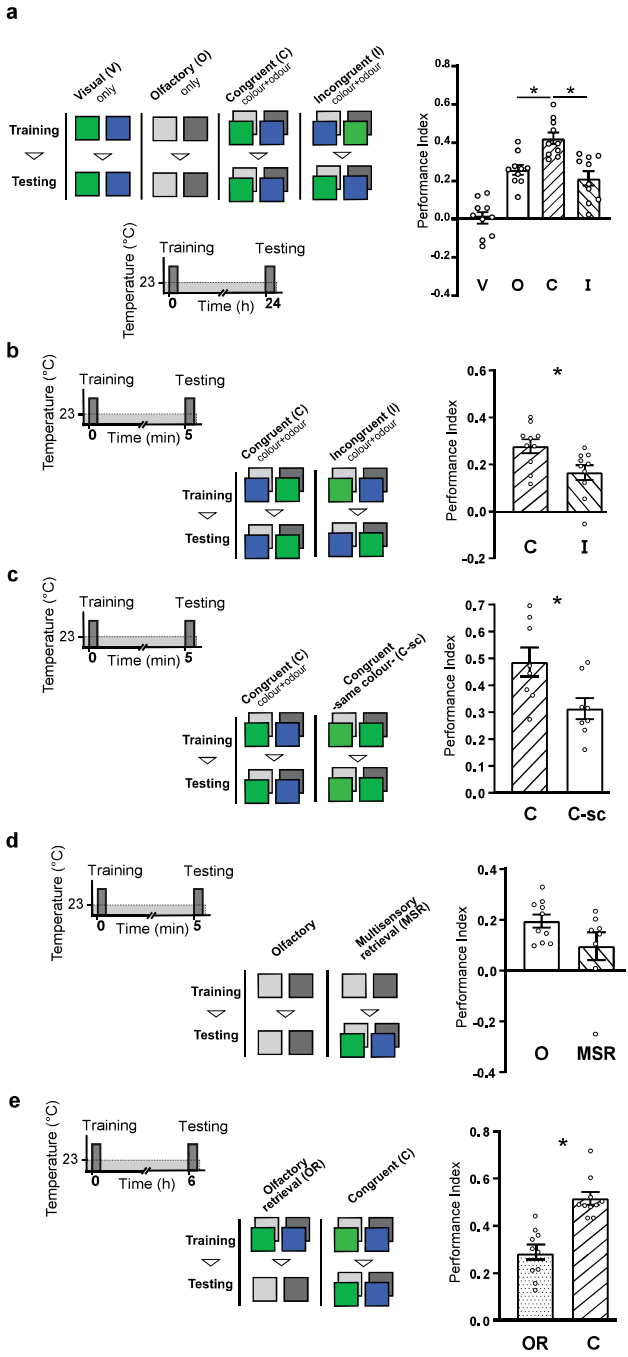

Figure 2

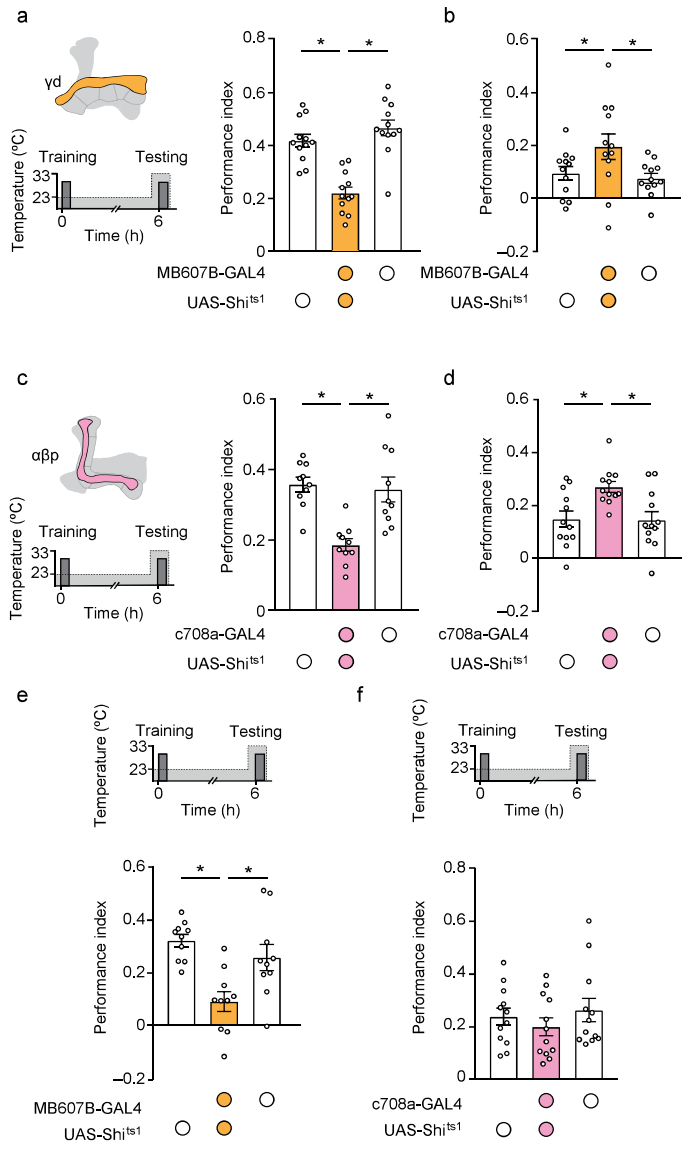

Figure 2 Supplement 1

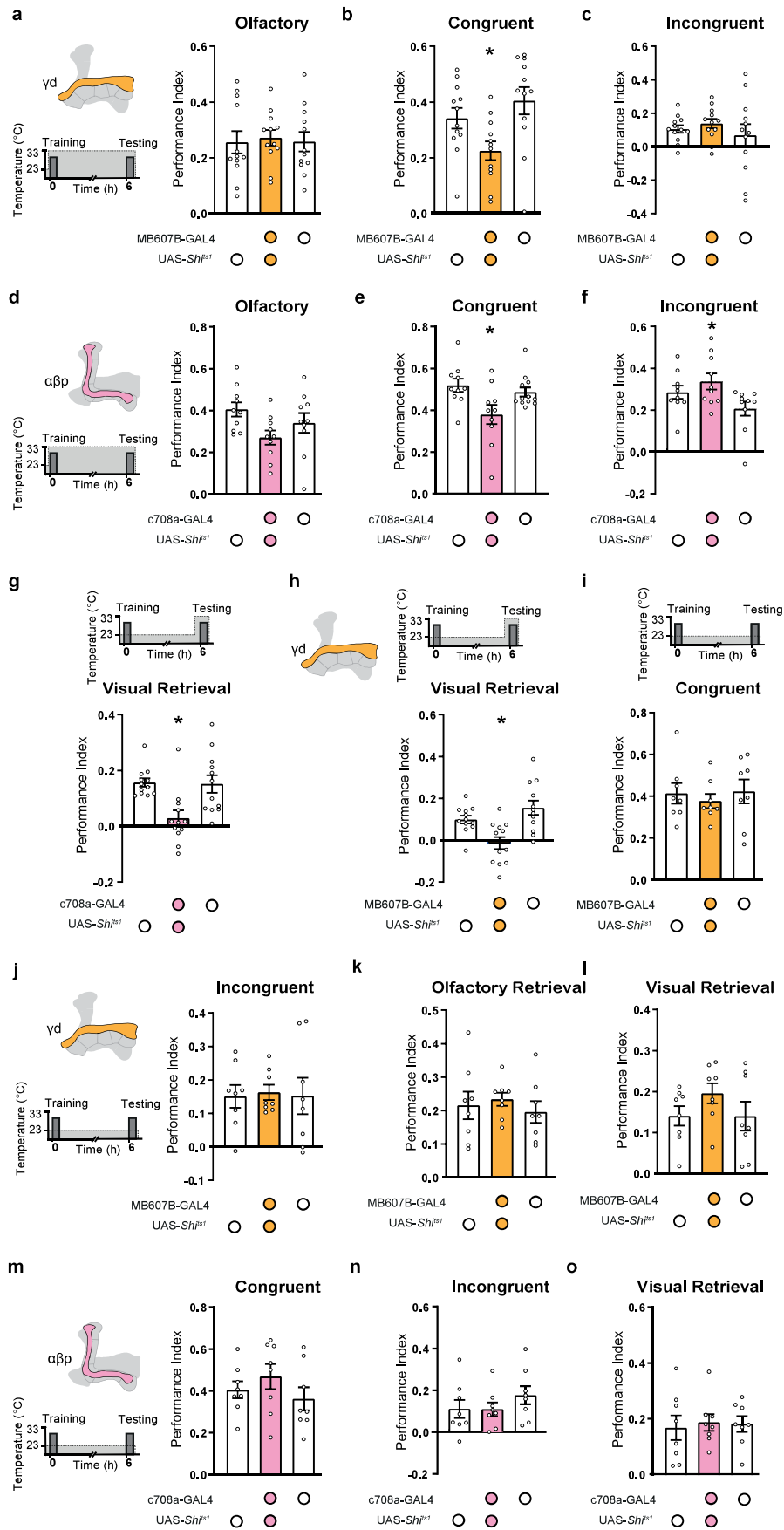

Figure 3

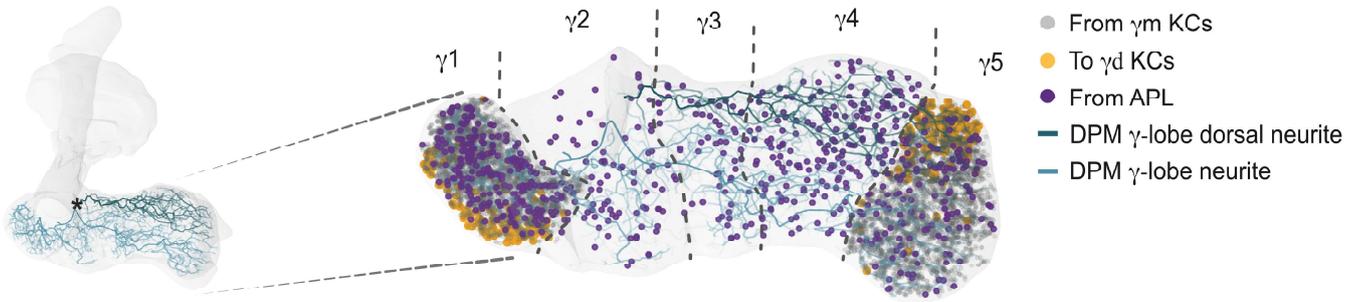

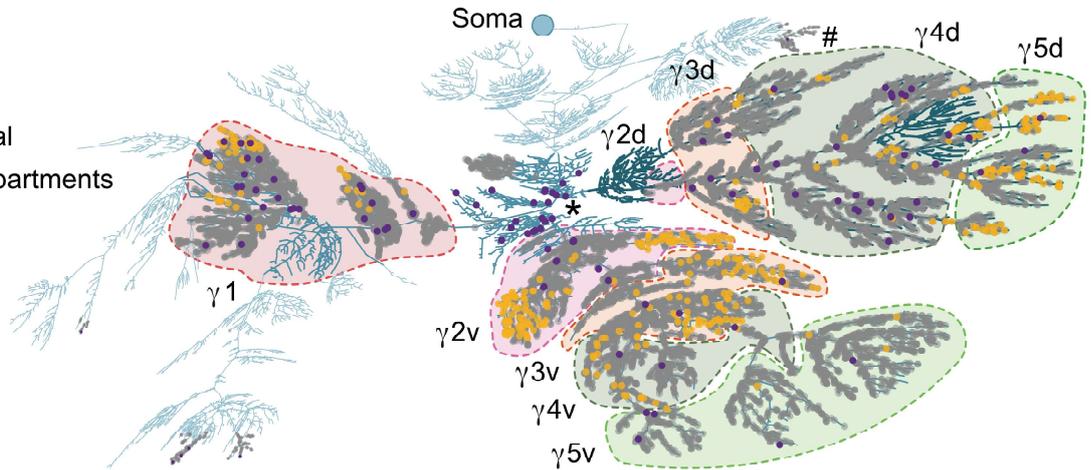

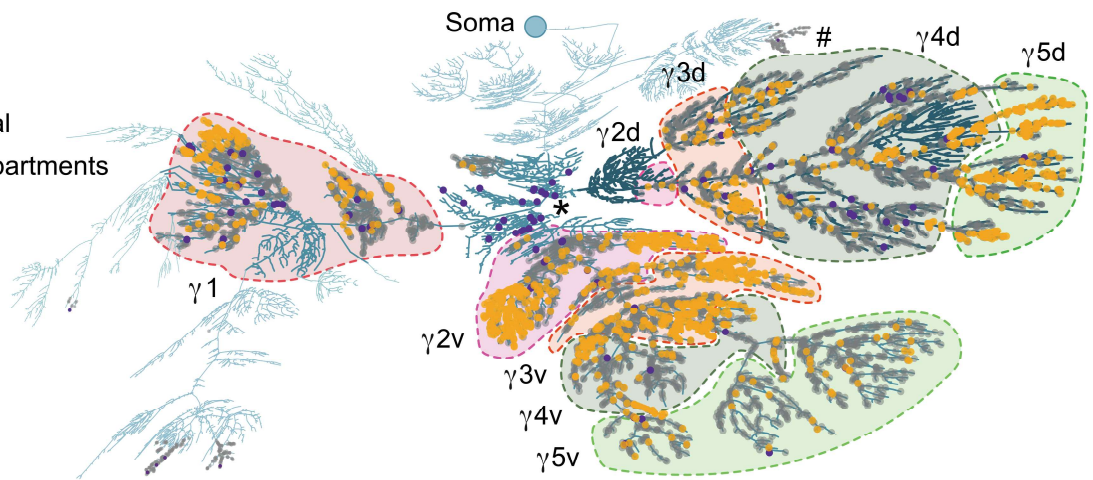

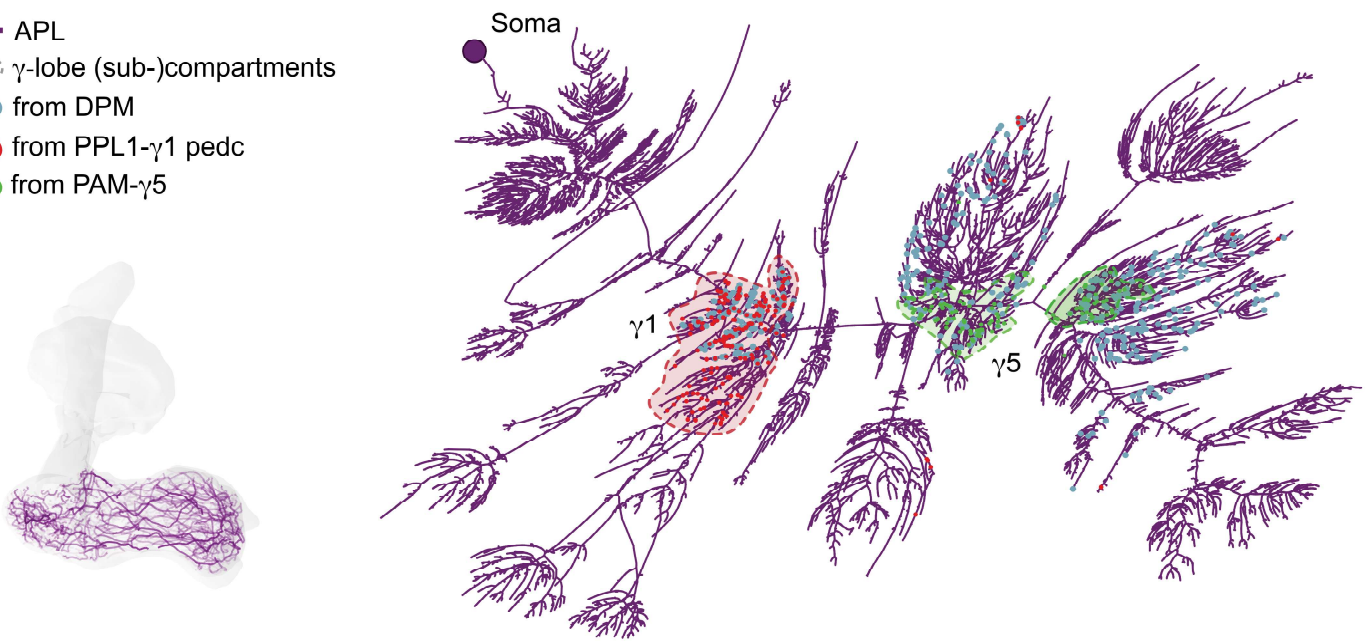

Figure 4

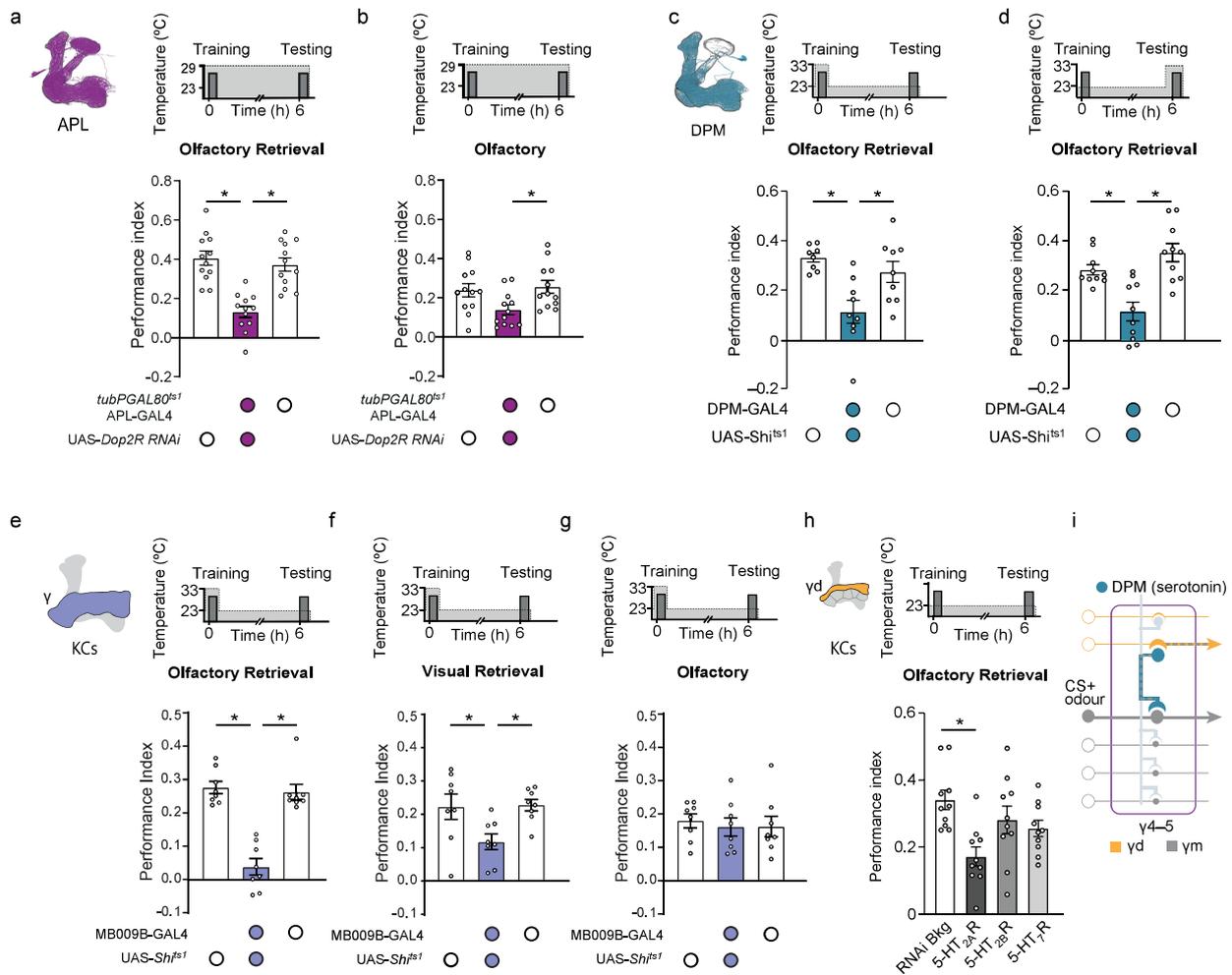

Figure 4 Supplement 1

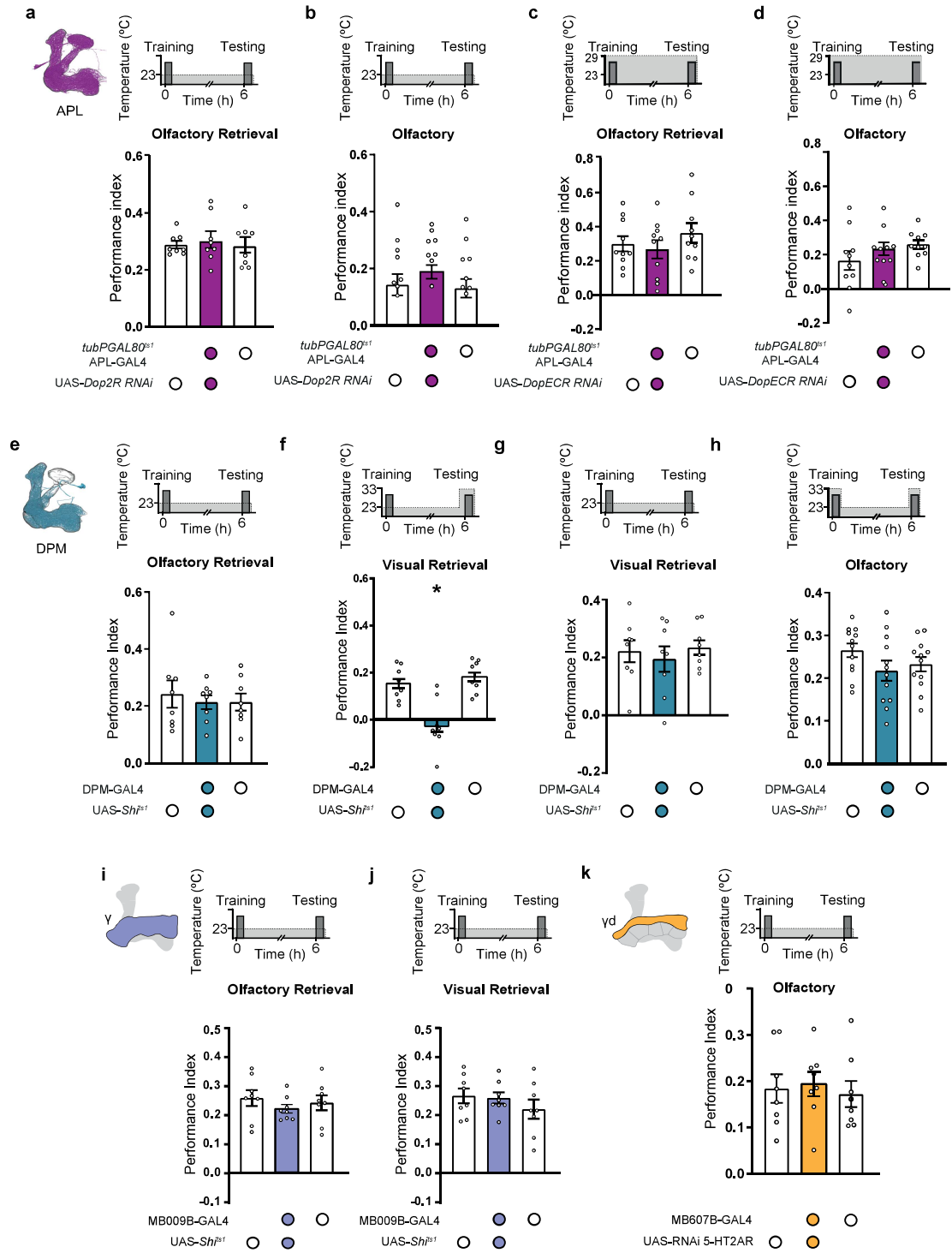

Figure 5

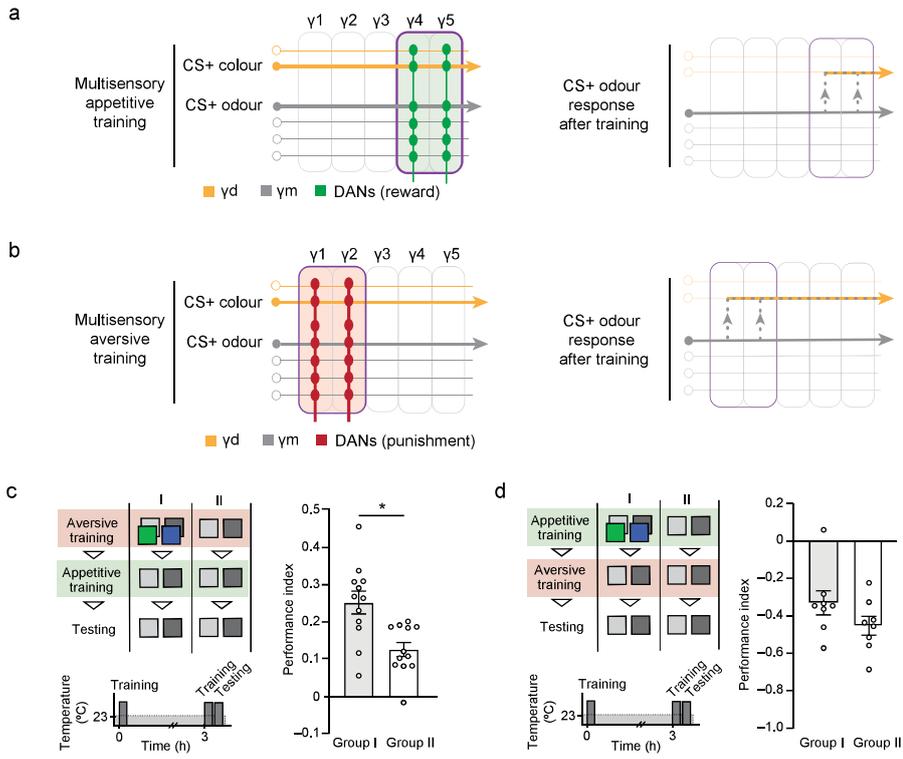

Figure 5 Supplement 1

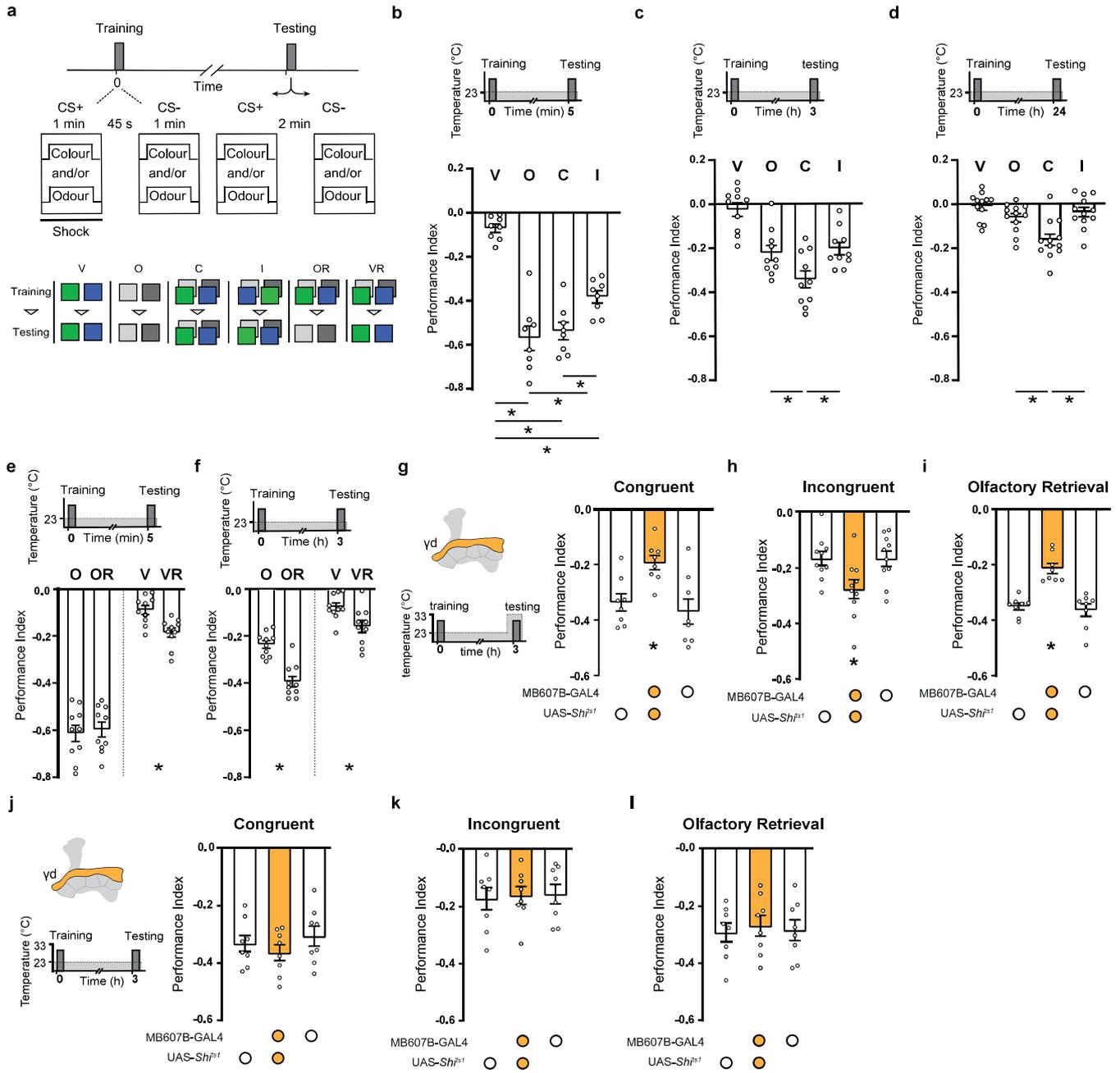